\documentclass[twoside,11pt]{8ssmmp} 

\usepackage{amsfonts}
\usepackage{amssymb}
\usepackage{amsmath}
\usepackage{fancyhdr}    
\usepackage{equations}
\usepackage{epic}
\usepackage{eepic}
\usepackage{epsf}
\usepackage{graphicx}
\usepackage{rotating}
\usepackage{hyperref}

\newcommand\rf[1]{(\ref{eq:#1})}
\newcommand\lab[1]{\label{eq:#1}}
\newcommand\nonu{\nonumber}
\newcommand\br{\begin{eqnarray}}
\newcommand\er{\end{eqnarray}}
\newcommand\be{\begin{equation}}
\newcommand\ee{\end{equation}}

\newcommand\lb{\lbrack}
\newcommand\rb{\rbrack}

\renewcommand\({\left(}
\renewcommand\){\right)}

\newcommand\bc{\begin{center}}
\newcommand\ec{\end{center}}

\newcommand\partder[2]{\frac{{\partial {#1}}}{{\partial {#2}}}}

\renewcommand\d{\delta}

\newcommand\eps{\epsilon}
\newcommand\vareps{\varepsilon}
\newcommand\g{\gamma}
\newcommand\G{\Gamma}

\newcommand\h{\frac{1}{2}}
\renewcommand\k{\kappa}
\renewcommand\l{\lambda}
\renewcommand\L{\Lambda}
\newcommand\m{\mu}
\newcommand\n{\nu}
\newcommand\om{\omega}

\renewcommand\P{\Phi}
\newcommand\pa{\partial}

\renewcommand\r{\rho}

\newcommand\z{\zeta}

\newcommand\cH{{\mathcal H}}

\newcommand{\ct}[1]{\cite{#1}}
\newcommand{\bib}[1]{\bibitem{#1}}

\newcommand\PRD[3]{\textsl{Phys. Rev.} \textbf{D#1} (#2) #3}

\newcommand\PLB[3]{\textsl{Phys. Lett.} \textbf{#1B} (#2) #3}
\newcommand\CQG[3]{\textsl{Class. Quantum Grav.} \textbf{#1} (#2) #3}

\newcommand\IJMPD[3]{\textsl{Int. J. Mod. Phys.} \textbf{D#1} (#2) #3}

\newcommand\MPLA[3]{\textsl{Mod. Phys. Lett.} \textbf{A#1} (#2) #3}

\newcommand\udot{\stackrel{.}{u}}

\newcommand\Bdot{\stackrel{.}{B}}
\newcommand\Hdot{\stackrel{.}{H}}

\font \msb=msbm10 scaled \magstep1
\newcommand{\IR}{\mbox{\msb R} }

\renewcommand{\headrulewidth}{0.4pt}

\begin{document}
 \baselineskip=11pt

\title{A New Venue of Spontaneous Supersymmetry Breaking in Supergravity
}
\author{\bf{Eduardo Guendelman and Mahary Vasihoun}
\hspace{.25mm}\thanks{\,e-mail address:
guendel@bgu.ac.il, mahary@bgu.ac.il}
\\ \normalsize{Physics Department, Ben Gurion University of the Negev}\\
\normalsize{Beer Sheva, Israel} \vspace{2mm} \\ 
\bf{Emil Nissimov and Svetlana Pacheva}\hspace{.25mm}\thanks{\,e-mail
address: nissimov@inrne.bas.bg, svetlana@inrne.bas.bg}
\\ \normalsize{Institute for Nuclear Research and Nuclear Energy}\\
\normalsize{Bulgarian Academy of Sciences, Sofia, Bulgaria} }

\date{}

\maketitle

\begin{abstract}
We present a qualitatively new mechanism for dynamical spontaneous breakdown of 
supersymmetry in supergravity. Specifically, we construct a modified formulation of 
standard minimal $N=1$ supergravity as well as of anti-de Sitter supergravity in terms 
of a non-Riemannian spacetime volume form (generally covariant integration
measure density). The new supergravity formalism naturally triggers the appearance of a 
{\em dynamically generated cosmological constant} as an arbitrary integration 
constant which signifies {\em spontaneous (dynamical) breaking of supersymmetry}.
Applying the new formalism to anti-de Sitter supergravity allows us to appropriately 
choose the above mentioned arbitrary integration constant so as to obtain simultaneously
a {\em very small effective observable cosmological constant} as well as a
{\em large physical gravitino mass} as required by modern cosmological scenarios for 
slowly expanding universe of the present epoch.
\end{abstract}

\section{Introduction}
\label{intro}

Supersymmetry is a fundamental extended space-time symmetry of Nature,
believed to manifest itself at ultra-high energies, which is unifying bosons 
(integer-spin particles) and fermions (half-integer spin particles). Among the
prinicipal theoretical highlights of supersymmetry one should mention:
drastically reducing the number of apriori independent physical
parameters; drastically reducing (in some cases even eliminating) ultraviolet 
divergencies in quantum field theories; providing possible solution of the 
hierachy/fine-tuning problems in high-energy elementary particles
phenomenology; naturally appearing within the context of modern string theory.
For a very short list of basic references see
\ct{susy-basic,weinberg-3,superstring,freedman-proeyen}. Apart
from elementary particle physics and cosmology, supersymmetry plays an
increasingly active role both as conceptional paradigm and as important
mathematical tool in various other principal areas of theoretical physics and
mathematics such as theoretical condensed matter \ct{susy-condmat} as well
as in the modern theory of integrable (soliton) systems \ct{susy-integrable}.

Unfortunately, supersymmetry is {\em not an exact} symmetry of Nature.
Otherwise, we would observe bosonic (integer-spin) counterparts with the
same masses of the fundamental fermionic particles -- protons, electrons, {\em etc.}. 
Therefore, supersymmetry must be {\em spontaneously broken}
\ct{susy-basic,weinberg-3}.

Under spontaneous symmetry breakdown the symmetry-generating charges in the
respective (quantum) field theories are conserved whereas the ground states
(``vacuums'') are {\em not} invariant under the symmetry transformations.
Spontaneous symmetry breakdown is always accompanied by the appearance of
certain mass (energy) scale of the breakdown.
Typically, the scale of spontaneous symmetry breaking is generated by the
appearance of non-zero vacuum expectation values of certain (quantum) fields
non-trivially transforming under the pertinent symmetry group. The basic 
example is the {\em Brout-Englert-Higgs} mechanism in the Standard
Model of particle physics (see \textsl{e.g.} Ref.\ct{weinberg-1-2}).

In minimal supergravity (supersymmetric generalizations of ordinary Einstein general
relativity without interactions with other matter fields  -- for a recent account of 
modern supergravity theories and notations, see Ref.\ct{freedman-proeyen})
there is another way to spontaneously break supersymmetry (supersymmetric
Brout-Englert-Higgs effect) 
-- via {\em dynamical generation of non-zero cosmological constant} 
\ct{sugra-ssb,deser-zumino-77}. In what follows we describe a new theoretical 
framework where the above scenario is explicitly realized in a natural way.

The main idea of our current approach comes from Refs.\ct{TMT-orig} 
(for recent developments, see Refs.\ct{TMT-recent}), where some of us
have proposed a new class of gravity-matter theories based on the idea that the 
action integral may contain several terms built with {\em different}
spacetime volume-forms (generally covariant integration measure densities).
Namely, apart from terms built with the standard Riemannian volume-form
given in terms of the square-root of the determinant of the pertinent
Riemannian metric, we may have additional term(s) constructed with one or more
\ct{emergent-quintess} {\em alternative non-Riemannian volume form(s) 
on the spacetime manifold} in terms of one or more auxiliary antisymmetric tensor gauge
field(s) of maximal rank completely independent of the metric.
The latter formalism has lead to various new interesting results
in all types of known generally coordinate-invariant theories:

\begin{itemize}
\item
(i) $D=4$-dimensional models of gravity and matter fields containing one term with 
an alternative non-Riemannian integration measure appear to be promising candidates 
for resolution of the dark energy and dark matter problems, the fifth force problem, 
and a natural mechanism for spontaneous breakdown of global Weyl-scale symmetry
\ct{TMT-orig}-\ct{TMT-recent}.
\item
(ii) Study of reparametrization invariant theories of extended objects 
(strings and branes) based on employing of a modified non-Riemannian 
world-sheet/world-volume integration measure \ct{mstring} leads to dynamically 
induced variable string/brane tension and to simple string models of non-abelian 
confinement.
\item
(iii) In Refs.\ct{emergent-quintess} a new class of extended gravity-matter
models is constructed, built in terms of two independent non-Riemannian volume-forms 
on the underlying spacetime manifold, producing interesting cosmological implications
relating inflationary and today's slowly accelerating phases of the universe.
\end{itemize}

The principal results described in the next sections are as follows:

(a) First, we briefly outline the main properties of a general class of 
gravity-matter models with one non-Riemannian volume-form term. Specifically we
provide a consistent canonical Hamiltonian analysis of the latter models exhibiting
the physical meaning of the pertinent auxiliary fields which are absent in
the standard formulation of gravity-matter actions in terms of the ordinary
Riemannian spacetime volume-form (see also second Ref.\ct{emergent-quintess}).

(b) The new non-Riemannian volume-form formalism is then applied to minimal $N=1$ 
supergravity in $D=4$-dimensional spacetime. This naturally triggers the appearance
of a {\em dynamically generated} cosmological constant as an arbitrary integration
constant, which signifies a new explicit mechanism of spontaneous (dynamical) 
breaking of supersymmetry.

(c) Applying the same formalism to {\em anti-de Sitter} supergravity allows us to 
appropriately choose the above mentioned arbitrary integration constant so as to 
obtain simultaneously a very small effective observable cosmological constant 
as well as a very large physical gravitino mass.

\section{Gravity-Matter Models With a Non-Riemannian Spacetime Volume-Form}
\label{TMMT}

Let us consider the following non-standard gravity-matter system of a general form
whose action is a linear combination (with $c_{1,2}$ -- some constants) of one term
with an alternative non-Riemannian spacetime volume-form and another one with the 
standard Riemannian one:
\be
S = c_1 \int d^D\!x \,\P (B) \Bigl\lb L^{(1)} + \P (H) \Bigr\rb
+ c_2 \int d^D\!x \sqrt{-g}\, L^{(2)}
\lab{TMT-gen}
\ee

Here the following notations are used:

\begin{itemize}
\item
The alternative volume element in the first term of \rf{TMT-gen} is given by 
the following {\em non-Riemannian} integration measure density:
\be
\Phi(B) \equiv \frac{1}{(D-1)!}\vareps^{\m_1\ldots\m_D}\, \pa_{\m_1} B_{\m_2\ldots\m_D}
\; ,
\lab{Phi-B}
\ee
where $B_{\m_1\ldots\m_{D-1}}$ is an auxiliary rank $(D-1)$ antisymmetric tensor gauge
field. 
The latter can also be parametrized in terms of $D$ auxiliary scalar fields 
$B_{\m_1\ldots\m_{D-1}}= \frac{1}{D}\vareps_{I J_1 \ldots J_{D-1}} 
\phi^{I} \pa_{\m_1} \phi^{J_1}\ldots \pa_{\m_{D-1}} \phi^{J_{D-1}}$,
so that: $\Phi(B) = \frac{1}{D!}\vareps^{\m_1\ldots\m_D}\,\vareps_{I_1 \ldots I_D} 
\pa_{\m_1} \phi^{I_1} \ldots \pa_{\m_D} \phi^{I_D}$, but we will stick to
the definition \rf{Phi-B}. The volume element in the second term of \rf{TMT-gen}
is given by the standard Riemannian integration measure density $\sqrt{-g}$,
where $g \equiv \det\Vert g_{\m\n}\Vert$ is the determinant of the corresponding
Riemannian metric $g_{\m\n}$.
\item
The Lagrangians $L^{(1,2)} \equiv \frac{1}{2\k^2} R + L^{(1,2)}_{\rm matter}$
include both standard Einstein-Hilbert gravity action as well as
matter/gauge-field parts. Here $R=g^{\m\n} R_{\m\n}(\G)$ is the scalar curvature 
within the first-order (Palatini) formalism and $R_{\m\n}(\G)$ is the Ricci tensor 
in terms of the independent affine connection $\G^\m_{\l\n}$.
\item
In general, the second Lagrangian $L^{(2)}$ might contain also higher curvature terms 
like $R^2$ (see \textsl{e.g.} first Ref.\ct{emergent-quintess}).
\item
In the first {\em modified-measure term} of the action \rf{TMT-gen}
we have included an additional term containing the field-strength $\P (H)$ of
another auxiliary rank $(D-1)$ antisymmetric tensor gauge field
$H_{\m_1\ldots\m_{D-1}}$:
\be
\Phi(H) \equiv \frac{1}{(D-1)!}\vareps^{\m_1\ldots\m_D}\, \pa_{\m_1} H_{\m_2\ldots\m_D}
\; ,
\lab{Phi-H}
\ee
whose presence is crucial for non-triviality of the model. 
Such term would be purely topological (total divergence) one if included in standard
Riemannian integration measure action like the second term with $L^{(2)}$ on the r.h.s.
of \rf{TMT-gen}.
\end{itemize}

The auxiliary gauge fields $B_{\m_1\ldots\m_{D-1}}$ and $H_{\m_1\ldots\m_{D-1}}$ 
turn out to be pure-gauge (non-propagating) degrees of freedom, however, both are
leaving remnants which play crucial role in the sequel (see also next Section).
Namely, varying \rf{TMT-gen} w.r.t. $H$ and $B$ tensor gauge fields we get:
\br
\pa_\m \Bigl(\frac{\Phi(B)}{\sqrt{-g}}\Bigr) = 0 \;\; \to \;\;
\frac{\Phi(B)}{\sqrt{-g}} \equiv \chi = {\rm const} \; ,
\lab{chi-const} \\
\pa_\m \Bigl\lb L^{(1)} + \Phi(H)\Bigr\rb = 0 \;\; \to \;\;
L^{(1)} + \Phi(H)
= M  = {\rm const} \; ,
\lab{L1-const}
\er
where $\chi$ (ratio of the two measure densities) and $M$ are {\em arbitrary 
integration constants}. The meaning of $\chi$ and $M$ from the point of view
of canonical Hamiltonian formalism is elucidated in the next section.

Now, varying \rf{TMT-gen} w.r.t. $g^{\m\n}$ and taking into account 
\rf{chi-const}--\rf{L1-const} we arrive at the following effective Einstein
equations (in the first-order formalism):
\be
R_{\m\n}(\G) - \h g_{\m\n} R + \L_{\rm eff} g_{\m\n} = \k^2 T^{\rm eff}_{\m\n},
\lab{einstein-eff}
\ee
with effective energy-momentum tensor:
\be
T^{\rm eff}_{\m\n} = g_{\m\n} L^{\rm eff}_{\rm matter} -
2\partder{L^{\rm eff}_{\rm matter}}{g^{\m\n}} \;\; , \;\;
L^{\rm eff}_{\rm matter} \equiv \frac{1}{c_1\chi + c_2} 
\Bigl\lb c_1 L^{(1)}_{\rm matter} + c_2 L^{(2)}_{\rm matter}\Bigr\rb \; ,
\lab{T-eff}
\ee
and with a {\em dynamically generated effective cosmological constant}
$\L_{\rm eff}$ thanks to the non-zero integration constants $M,\,\chi$: 
\be
\L_{\rm eff} = \k^2 \(c_1\chi + c_2\)^{-1}\,\chi M  \; .
\lab{CC-eff}
\ee

\section{Canonical Hamiltonian Treatment}
\label{hamiltonian}

In what follows we restrict our attention to $D=4$-dimensional spacetime.
For convenience we will introduce the following short-hand notations for the
field-strengths \rf{Phi-B} and \rf{Phi-H} of the auxiliary 3-index antisymmetric 
gauge fields $B_{\m\n\l},\, H_{\m\n\l}$ (the dot indicating time-derivative): 
\br
\P (B) = \Bdot + \pa_i B^i \quad, \quad 
B = \frac{1}{3!} \vareps^{ijk} B_{ijk} \;\; ,\;\;
B^i = - \h \vareps^{ijk} B_{0jk} \; ,
\lab{B-can} \\
\P (H) = \Hdot + \pa_i H^i \quad, \quad 
H = \frac{1}{3!} \vareps^{ijk} H_{ijk} \;\; ,\;\;
H^i = - \h \vareps^{ijk} H_{0jk} \; ,
\lab{H-can}
\er
According to the general form of the action \rf{TMT-gen} (for simplicity we
set here $c_{1,2}=1$) the pertinent canonically conjugated momenta read:
\br
\pi_B = L^{(1)} (u,\udot) + \frac{1}{\sqrt{-g}}(\Hdot + \pa_i H^i) \; ,
\nonu \\
\pi_H = \frac{1}{\sqrt{-g}}(\Bdot + \pa_i B^i) \; ,
\lab{can-momenta-aux}
\er
where $(u,\udot)$ collectively denote the set of the basic gravity-matter
canonical variables ($(u)=(g_{\m\n},\,{\rm matter~fields})$) and their velocities, 
and:
\be
\pi_{B^i} = 0 \quad,\quad \pi_{H^i} = 0 \; .
\lab{can-momenta-zero}
\ee
Eqs.\rf{can-momenta-zero} imply that $B^i, H^i$ will in fact appear as Lagrange 
multipliers for certain first-class Hamiltonian constraints 
(see Eqs.\rf{pi-B-pi-H-const} below). 

Using \rf{can-momenta-aux}, for the canonical
momenta conjugated to the basic gravity-matter canonical variables we obtain: 
\be
p_u = \pi_H \frac{\pa}{\pa \udot} \Bigl(\sqrt{-g} L^{(1)} (u,\udot)\Bigr) + 
\frac{\pa}{\pa \udot} \Bigl(\sqrt{-g} L^{(2)} (u,\udot) \Bigr)\; .
\lab{can-momenta-u}
\ee
Now, from \rf{can-momenta-aux} and \rf{can-momenta-u} we obtain the velocities 
$\Hdot=\Hdot\bigl(u,p_u,\pi_H,\pi_B\bigr)$, $\Bdot=\Bdot(u,\pi_H)$ and 
$\udot=\udot\bigl(u,p_u,\pi_H,\pi_B\bigr)$ as functions of the respective canonically
conjugate momenta, wherefrom the canonical Hamiltonian corresponding to \rf{TMT-gen}:
\br
\cH = p_u \udot  + \pi_B \Bdot + \pi_H \Hdot \phantom{aaaaaaaaaaaaa}
\nonu \\
- (\Bdot + \pa_i B^i) \Bigl\lb L^{(1)}(u,\udot) + 
\frac{1}{\sqrt{-g}}(\Hdot + \pa_i H^i) \Bigr\rb - {\sqrt{-g}} L^{(2)}(u,\udot)
\lab{can-hamiltonian}
\er
acquires the following form as function of the canonically conjugated
variables (here $\udot = \udot \bigl(u,p_u,\pi_H,,\pi_B\bigr)$:
\br
\cH = p_u \udot - \pi_H \sqrt{-g} L^{(1)} (u,\udot) - \sqrt{-g} L^{(2)} (u,\udot)
\nonu \\
+ \sqrt{-g} \pi_H \pi_B - \pa_i B^i \pi_B - \pa_i H^i \pi_H \; .
\lab{can-hamiltonian-final}
\er
From \rf{can-hamiltonian-final} we deduce that indeed $B^i, H^i$ are Lagrange 
multipliers for the first-class Hamiltonian constraints:
\be
\pi_H = \chi = {\rm const} \quad ,\quad \pi_B = M = {\rm const} \; ,
\lab{pi-B-pi-H-const}
\ee
which (in virtue of \rf{can-momenta-aux}) are the canonical Hamiltonian counterparts 
of Lagrangian constraint equations of motion \rf{chi-const}-\rf{L1-const}.

We conclude that the canonical Hamiltonian treatment of \rf{TMT-gen} reveals the meaning
of the auxiliary 3-index antisymmetric tensor gauge fields
$B_{\m\n\l},\, H_{\m\n\l}$. Namely, the canonical momenta $\pi_B,\,\pi_H$ 
conjugated to the ``magnetic'' parts $B,H$ \rf{B-can}-\rf{H-can}
of the respective  tensor gauge fields are constrained through Dirac first-class 
constraints \rf{pi-B-pi-H-const} to be constants identified with the arbitrary 
integration constants $\chi,\, M$ \rf{chi-const}-\rf{L1-const} arising within the 
Lagrangian formulation of the model. The canonical momenta 
$\pi_B^i,\,\pi_H^i$ conjugated to the ``electric'' parts $B^i,H^i$ 
\rf{B-can}-\rf{H-can} of the auxiliary 3-index antisymmetric tensor gauge field
are vanishing \rf{can-momenta-zero} which makes the latter canonical Lagrange 
multipliers for the above Dirac first-class constraints.

\section{Supersymmetric Brout-Englert-Higgs Effect in Minimal Supergravity}
\label{susy-ssb}

Let us now apply the above formalism to construct a non-Riemannian spacetime
volume-form version of simplest $N=1$ supergravity in $D=4$.

Let us recall the standard component-field action of $D=4$ minimal $N=1$ supergravity
(for definitions and notations we follow \ct{freedman-proeyen}):
\br
S_{\rm SG} = \frac{1}{2\k^2} \int d^4 x\, e
\Bigl\lb R(\om,e) - {\bar\psi}_\m \g^{\m\n\l} D_\n \psi_\l \Bigr\rb \; ,
\lab{SG-action} \\
e = \det\Vert e^a_\m \Vert \;\; ,\;\;
R(\om,e) = e^{a\m} e^{b\n} R_{ab\m\n}(\om) \; .
\lab{curv-scalar} \\
R_{ab\m\n}(\om) = \pa_\m \om_{\n ab} - \pa_\n \om_{\m ab}
+  \om_{\m a}^c \om_{\n cb} - \om_{\n a}^c \om_{\m cb} \; .
\lab{curvature} \\
D_\n \psi_\l = \pa_\n \psi_\l + \frac{1}{4}\om_{\n ab}\g^{ab}\psi_\l \;\; ,\;\;
\g^{\m\n\l} = e^\m_a e^\n_b e^\l_c \g^{abc} \; ,
\lab{D-covariant}
\er 
where all objects belong to the first-order ``vierbein'' (frame-bundle) formalism.

The vierbeins $e^a_\m$ (describing the graviton) and the 
spin-connection $\om_{\m ab}$ ($SO(1,3)$ gauge field acting on the gravitino
$\psi_\m$) are \textsl{a priori} independent fields (their relation arises
subsequently on-shell); $\g^{ab} \equiv \h \(\g^a \g^b - \g^b \g^a\)$
\textsl{etc.} denote antisymmetrized products of gamma-matrices with $\g^a$
being the ordinary Dirac gamma-matrices.
The invariance of the  action \rf{SG-action} under local supersymmetry 
transformations: 
\be
\d_\eps e^a_\m = \h {\bar\vareps}\g^a \psi_\m \;\; ,\;\;
\d_\eps \psi_\m = D_\m \vareps
\lab{local-susy}
\ee
follows from the invariance of the pertinent Lagrangian density up to a
total derivative: 
\be
\d_\eps \Bigl( e \bigl\lb R(\om,e) 
- {\bar\psi}_\m \g^{\m\n\l} D_\n \psi_\l \bigr\rb\Bigr) = 
\pa_\m \bigl\lb e\bigl({\bar\vareps}\z^\m\bigr)\bigr\rb \; ,
\lab{local-susy-L}
\ee
where $\z^\m$ functionally depends on the gravitino field $\psi_\m$.

We now propose a modification of \rf{SG-action} by replacing the standard
generally-covariant measure density $e = \sqrt{-g}$ by the alternative measure
density $\P(B)$ (Eq.\rf{Phi-B} for $D=4$):
\be
\Phi(B) \equiv \frac{1}{3!}\vareps^{\m\n\k\l}\, \pa_\m B_{\n\k\l} \; ,
\lab{Phi-4}
\ee
and we will use the general framework described above.
The modified supergravity action reads:
\be
S_{\rm mSG} = \frac{1}{2\k^2} \int d^4 x\, \Phi(B)\,
\Bigl\lb R(\om,e) - {\bar\psi}_\m \g^{\m\n\l} D_\n \psi_\l 
+ \frac{\vareps^{\m\n\k\l}}{3!\,e}\, \pa_\m H_{\n\k\l} \Bigr\rb \; ,
\lab{mSG-action}
\ee
where a new term containing the field-strength (Eq.\rf{Phi-H} for $D=4$) of a
3-index antisymmetric tensor gauge field $H_{\n\k\l}$ has been added.

The equations of motion w.r.t. $H_{\n\k\l}$ and $B_{\n\k\l}$ yield: 
\br
\pa_\m \Bigl(\frac{\Phi(B)}{e}\Bigr) = 0 \;\; \to \;\;
\frac{\Phi(B)}{e} \equiv \chi = {\rm const} \; ,
\lab{chi-const-1} \\
R(\om,e) - {\bar\psi}_\m \g^{\m\n\l} D_\n \psi_\l
+ \frac{\vareps^{\m\n\k\l}}{3!\, e}\, \pa_\m H_{\n\k\l} = 2 M \; ,
\lab{L-const}
\er
where $\chi$ and $M$ are arbitrary integration constants.

The action \rf{mSG-action} is invariant under local supersymmetry transformations
\rf{local-susy} supplemented by transformation laws for $H_{\m\n\l}$ and $\Phi(B)$:
\be
\d_\eps H_{\m\n\l} = - e\,\vareps_{\m\n\l\k}\bigl({\bar\vareps}\z^\k\bigr)
\quad, \quad \d_\eps \Phi(B) = \frac{\Phi(B)}{e}\,\d_\eps e \; ,
\lab{local-susy-H-Phi}
\ee
which algebraically close on-shell, 
\textsl{i.e.}, when Eq.\rf{chi-const-1} is imposed.

The appearance of the integration constant $M$ represents a {\em dynamically 
generated cosmological constant} in the pertinent gravitational equations of motion and, 
thus, it signifies a {\em spontaneous (dynamical) breaking of supersymmetry}. 

Indeed, varying \rf{mSG-action} w.r.t. $e^a_\m$:
\br
e^{b\n}R^a_{b\m\n} - \h {\bar\psi}_\m \g^{a\n\l} D_\n \psi_\l
+ \h {\bar\psi}_\n \g^{a\n\l} D_\m \psi_\l 
\nonu \\
+ \h {\bar\psi}_\l \g^{a\n\l} D_\n \psi_\m
+ \frac{e^a_\m}{2}\,\frac{\vareps^{\r\n\k\l}}{3!\, e}\, \pa_\r H_{\n\k\l} = 0
\lab{grav-eqs}
\er
and using Eq.\rf{L-const} (containing the arbitrary integration constant $M$) to 
replace the last $H$-term on the l.h.s. of \rf{grav-eqs}, the results is as
follows: we obtain the vierbein counterparts of the Einstein equations including a
dynamically generated {\em floating} cosmological constant term $e^a_\m M$:
\br
e^{b\n}R^a_{b\m\n} -\h e^a_\m R(\om,e) +  e^a_\m M = \k^2 T^a_\m \; ,
\phantom{aaaaaaaaaaaaaaaaaaa}
\lab{einstein-eqs} \\
\k^2 T^a_\m \equiv \h {\bar\psi}_\m \g^{a\n\l} D_\n \psi_\l
-\h e^a_\m {\bar\psi}_\r \g^{\r\n\l} D_\n \psi_\l
- \h {\bar\psi}_\n \g^{a\n\l} D_\m \psi_\l
- \h {\bar\psi}_\l \g^{a\n\l} D_\n \psi_\m \; .
\nonu
\er
Let us recall at this point that according to the classic paper \ct{deser-zumino-77} 
the sole appearance of a cosmological constant in supergravity, even in the absence 
of a manifest mass term for the gravitino, implies that the gravitino becomes 
{\em massive}, i.e., it absorbs the Goldstone fermion of spontaneous supersymmetry 
breakdown -- a {\em supersymmetric Brout-englert-Higgs effect}.

A significantly more interesting scenario occurs when applying the above
formalism with non-Riemannian spacetime volume-forms to anti-de Sitter (AdS) 
supergravity. Namely, let us start with the standard AdS supergravity action
(see \textsl{e.g.} Ref.\ct{freedman-proeyen}):
\br
S_{\rm AdS-SG} = \frac{1}{2\k^2} \int d^4 x\, e
\Bigl\lb R(\om,e) - {\bar\psi}_\m \g^{\m\n\l} D_\n \psi_\l 
- m\,{\bar\psi}_\m \g^{\m\n} \psi_\n - 2 \L_0 \Bigr\rb \; , \quad 
\lab{AdS-SG-action} \\
m \equiv \frac{1}{L} \quad ,\quad \L_0 \equiv - \frac{3}{L^2} \; .
\phantom{aaaaaaaaaaaaaaaaaaa}
\lab{mass-CC}
\er
The action \rf{AdS-SG-action} contains additional explicit mass term for the gravitino 
as well as a bare cosmological constant $\L_0$ balanced in a precise way $|\L_0|=3m^2$
so as to maintain local supersymmetry invariance and, in particular, keeping
the {\em physical gravitino mass zero} in spite of the presence of a ``bare''
gravitino mass term!

At this point let us stress that here we have AdS spacetime as a background
(``vacuum'') with curvature radius $L$ (unlike Minkowski background in the 
absence of a bare cosmological constant). Therefore, the notions of ``mass'' and
``spin'' are now given in terms of the eigenvalues of the Casimirs of the
unitary irreducible representations (discrete series) of the 
group of motion of AdS space $SO(2,3) \sim Sp(4,\IR)$ (for $D=4$) instead of the 
ordinary Poincare group ($SO(1,3) \ltimes R^4$) Casimirs (see \textsl{e.g.}
Ref.\ct{gazeau-novello}). Thus, on AdS background it is possible for the
gravitino to be massive even in the absence of a ``bare'' gravitino mass
term and, \textsl{vice-versa}, it will be massless even in the presence of a
large ``bare'' mass provided it is tuned up w.r.t. AdS cosmological constant
as in \rf{mass-CC}.

Now, following the same steps as with \rf{mSG-action} we construct a modified
AdS supergravity with non-Riemannian spacetime volume element:
\br
S_{\rm mod-AdS-SG} = \frac{1}{2\k^2} \int d^4 x\,\Phi(B)
\Bigl\lb R(\om,e) - {\bar\psi}_\m \g^{\m\n\l} D_\n \psi_\l 
\nonu \\
- m\,{\bar\psi}_\m \g^{\m\n} \psi_\n - 2 \L_0 + 
\frac{\vareps^{\m\n\k\l}}{3!\,e}\, \pa_\m H_{\n\k\l}\Bigr\rb \; ,
\lab{mod-AdS-SG-action}
\er
with $\Phi(B)$ as in \rf{Phi-4} and $m,\L_0$ as in \rf{AdS-SG-action}.
The action \rf{mod-AdS-SG-action} is invariant under local supersymmetry
transformations:
\br
\d_\eps e^a_\m = \h {\bar\vareps}\g^a \psi_\m \quad ,\quad
\d_\eps \psi_\m = \Bigl( D_\m - \frac{1}{2L}\g_\m \Bigr)\vareps \; ,
\nonu \\
\d_\eps H_{\m\n\l} = - e\,\vareps_{\m\n\l\k}\bigl({\bar\vareps}\z^\k\bigr)
\quad, \quad \d_\eps \Phi(B) = \frac{\Phi(B)}{e}\,\d_\eps e \; .
\er

The modified AdS supergravity action \rf{mod-AdS-SG-action}
will trigger dynamical spontaneous supersymmetry breaking resulting in the appearance 
of the dynamically generated floating cosmological constant $M$ as in 
Eq.\rf{einstein-eqs} which will add to the bare cosmological constant $\L_0$. 
Now we can use the freedom in
choosing the value of the \textsl{a priori} arbitrary integration constant
$M$ in order to match two basic requirements by modern cosmological scenarios for
slowly expanding universe of today \ct{slow-accel}.
Namely, we can achieve via appropriate choice of $M \simeq |\L_0|=3m^2$ a 
{\em very small effective observable cosmological constant}:
\be
\L_{\rm eff} = M + \L_0 = M - 3m^2 \ll |\L_0|
\lab{CC-eff-AdS-SG}
\ee
and, simultaneously, a {\em large physical gravitino mass} $m_{\rm eff}$:
\be
m_{\rm eff} \simeq m = \sqrt{\frac{1}{3}|\L_0|} \; ,
\lab{m-grav-AdS}
\ee
which will be very close to the large ``bare'' gravitino mass parameter 
$m = \sqrt{|\L_0|/3}$ since now because of the smallness of $\L_{\rm eff}$
\rf{CC-eff-AdS-SG} the background spacetime geometry becomes almost flat.

\section{Conclusions}
\label{conclude}

We have shown that applying the formalism with non-Riemannian spacetime
volume-forms in gravity/matter theories provides a simple mechanism for
dynamical generation of a cosmological constant. The latter appears as a
conserved Dirac-constrained canonical momentum conjugated to the auxiliary
maximal-rank antisymmetric gauge field building up the non-Riemannian
volume-form. In the context of modified
minimal $N=1$ supergravity defined in terms of a non-Riemannian spacetime
volume-form the dynamically generated cosmological constant triggers
spontaneous supersymmetry breakdown and gravitino mass generation
(supersymmetric Brout-Englert-Higgs effect). Upon constructing  modified 
anti-de Sitter supergravity with a non-Riemannian spacetime volume-form we can 
fine-tune the dynamically generated cosmological integration constant in order 
to achieve simultaneously a very small physical observable cosmological constant 
and a very large physical observable gravitino mass -- a paradigm of modern 
cosmological scenarios for slowly expanding universe of the present epoch.

\section*{Acknowledgments.} 
E.G. and E.N. are sincerely thankful to Prof. Branko Dragovich and the organizers
of the {\em Eight Meeting in Modern Mathematical Physics} in Belgrade for cordial 
hospitality. We gratefully acknowledge support of our collaboration through the 
academic exchange agreement between the Ben-Gurion University and the Bulgarian 
Academy of Sciences.
E.N. and S.P. are partially supported by Bulgarian NSF Grant \textsl{DFNI T02/6}.
S.P. has received partial support from European COST action MP-1210.


\end{document}